\documentstyle[12pt,epsfig,lathuile]{article}
\begin{document}
%
\title{
LINEAR COLLIDER PHYSICS\protect\thanks{~~Invited talk at Les Rencontres de Physique 
de la Vallée d'Aoste, La Thuile, Italy; February 27-March 4, 2000}}
\author{David J. Miller \\
{\em Physics and Astronomy, University College London,} \\ {\em Gower St., London WC1E
6BT, UK.  dave.miller@cern.ch}}

\maketitle 
\baselineskip=14.5pt 
\begin{abstract} 
Studies of the physics potential of the Future Linear Collider are establishing 
a broad programme which will start in the region of 350 to 500 
GeV C. of M. energy.  The main goal is to understand why the standard model
works; by studying the properties of the Higgs sector, if it is within reach,
and by exploring the complex world of Supersymmetry, if it is real.  If the Higgs boson is not
found soon, then the Linear Collider can test the standard model with high
precision measurements, both at 
energies approaching 1 TeV and with high statistics at the $Z^0$.
\end{abstract} 

\baselineskip=17pt 
\newpage 

\section{Where are we?} 

If there is a decision in 2003 to go ahead with a Linear Collider
programme  the machine and detectors could be ready for operation in 2010.   By
then the LHC at CERN will have overcome any teething troubles and be well into
its mainstream  programme.  In particular, it may have found at least one Higgs
boson, if the Higgs mass lies in the  range below 250 GeV which is favoured by
current fits to the standard model~\cite{lepewwg},  or if Higgs bosons are
generated by SUSY theories which are perturbative up to a high
scale~\cite{espinosa}.  There may  still be discovery opportunities for the
Linear Collider, but its main role will be to make clean precision
measurements, whether or not new physics  has already been found.  These will
either identify just what it is that has been found and see how it explains
electroweak symmetry breaking, or they will further constrain the standard
model in order to anticipate how the symmetry will be broken.     

Studies of physics for the future linear collider are now focussing on the case
to be made for funding.  This case will be presented in Europe in Spring 2001
and at about the same time in the USA. Japan has plans on a similar timescale.
By 1 March 2001 DESY will produce a 
Technical Design Report (TDR)  for the TESLA machine, including an X-ray 
free-electron-laser  facility for biomedical and condensed matter studies. 
European physicists are collaborating in the  2nd  ECFA/DESY Study~\cite{ecfad}
which will provide  the sections of the TDR on the physics  programme, the
detector and the machine-detector interface.  Unlike earlier studies, this TDR
will  contain a full costing of all parts of the programme.  
The ECFA/DESY Study has held
a series of workshops in Orsay,  Lund, Frascati, Oxford and Obernai (France),
continuing in Padova in May 2000 and DESY in  September 2000.  There will be a
TESLA ``launch conference'' in Autumn 2001 when the TDR is ready.   In parallel
with the ECFA/DESY study there are North American workshops~\cite{charlie}, 
the next at LBL in
March  2000~\cite{LBL2000}, ACFA workshops in Asia~\cite{acfa} and a 
worldwide series where the three
regions exchange ideas.  The  next worldwide Linear Collider Workshop (LCWS)
will be at Fermilab in October 2000.  (Previous  
LCWS were at Saariselka,
Finland~\cite{saariselka}; Waikoloa, Hawaii~\cite{waikoloa}; Morioka-Appi, 
Japan~\cite{morioka}; Sitges, Spain~\cite{sitges})

There is no space here for a description of the possible detector to do this
physics.  The ECFA/DESY  study has based most of its simulations on the
detector design outlined in the 1997 Conceptual Design  Report~\cite{cdr},
though the version now being used includes developments in vertexing - which 
will improve the identification of charm, a higher magnetic field - which
will improve the  resolution on missing masses (see section \ref{mm} below), and
more finely segmented calorimetry -  expressly driven by the need to make
the best possible measurements of hadron jets  in multi jet final  states.  

\section{Requirements on the Accelerator}

\begin{figure}[htbp]
         \centering
         \epsfxsize=10cm
         \epsffile{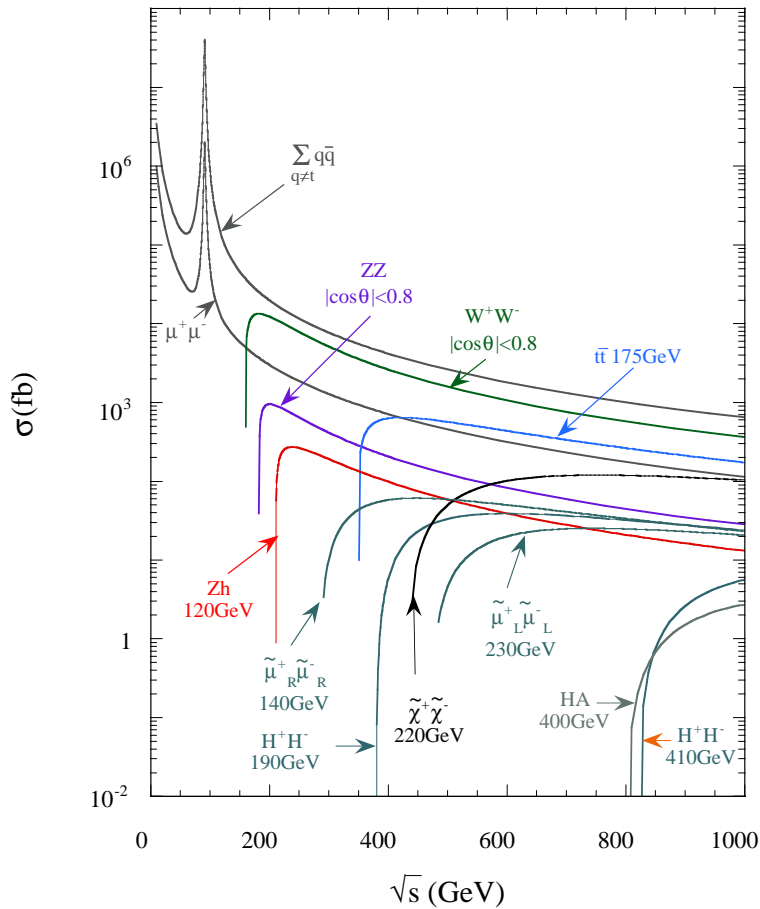}
 \caption{\it
      Cross sections for real and possible processes at a linear collider.
    \label{crosx} }
\end{figure}
The main factor which has delayed Linear Collider development has been the
imperative need to  achieve very high luminosities, much higher than  LEP which
delivers $250 pb^-1$/year at best.  LEP  felt like a high luminosity machine
when it was sitting on 30 nb total cross section at the $Z^0$  resonance peak,
but even with improved luminosity at LEP2 we can still only produce handfuls of
$Z^0$-pair events.  Figure~\ref{crosx} shows how the cross sections for some
interesting processes should  vary with beam energy.  The standard reference
process is $e^+ e^-  \rightarrow \mu^+ \mu^-$, ``1 unit  of R'', whose cross
section falls like 1/s, as do $t \bar{t}$ or light Higgs boson  production in
the  higgstrahlung channel $e^+ e^-  \rightarrow Z^0 H$.  Some other channels, with
dominant t-channel  exchanges, rise more slowly from their thresholds - for
example single W production $e^+ e^-   \rightarrow e^{\pm} \nu W^{\mp}$, $e^+
e^-  \rightarrow \nu \bar{\nu} H$ or $e^+ e^-  \rightarrow  \nu \bar{\nu} W^+
W^- $ or some of the SUSY processes in Figure~\ref{crosx}.  

Table~\ref{tab:rates} shows the numbers of events to be expected in some interesting channels
on the assumption  that $500fb^{-1}$ can be collected at $500 GeV$, or $1000
fb^{-1}$ at $1 TeV$.  The  relativistic
shrinkage of emittance in the machine will actually help to increase the
luminosity at higher  energies, so long as the beam alignment can be kept under
control.  With event numbers like these  definitive studies can be made of, for
instance, the branching ratios of a light Higgs boson - see below.
\begin{table}[b]
\centering
\begin{tabular}{|lc|lc|}  \hline
\multicolumn{2}{|c|}{$500 fb^{-1}$} & 
\multicolumn{2}{c|}{$1000 fb^{-1}$} \\
\multicolumn{2}{|c|}{@$\sqrt{s}=500GeV$} &
\multicolumn{2}{c|}{@$\sqrt{s}=1000GeV$} \\ \hline
$30,000$ & $ZH_{120}$ & $3,000$ & $\nu \bar{\nu}H_{500}$ \\
$50,000$ & $\nu \bar{\nu}H_{120}$ & $2,000$ & $WW \nu \bar{\nu}$, No Higgs \\
$3.5 \times 10^6$ & $WW$ or $e \nu W$ & $6,000$ & $WWZ$, No Higgs \\ \hline
\end{tabular}
\caption{Numbers of events in some possible channels.} \label{tab:rates}
\end{table}

To get $500 fb^{-1}$ in a $10^7$sec machine-year will require a luminosity of
\\
$5.10^{34} cm^{- 2} sec^{-1}$.  We believe this to be possible, both because of
what has been learned from running the Stanford Linear Collider, and because of the
advances beyond that made by the competing linear  collider development teams -
including the Russians until they were forced by financial circumstances  to
drop out.  The single most important advance over SLC will be a reduction in
vertical spot size  from a few microns to a few tens of nanometres.  A big part
of the technology needed for this was  proven by the world wide collaboration
that worked with the Final Focus Test Beam at SLC.  Another  big part of it has
to come from high brightness electron guns and from damping rings to reduce
the  beam emittance.  The most advanced damping ring test is under way at KEK
in Japan.  

The largest part  of the cost will be for the accelerating structure
of the linac itself.  Here the competition is  primarily between the
normally-conducting  X-band structure being developed by SLAC~\cite{nlc} 
and KEK~\cite{jlc} and  the
superconducting TESLA concept from DESY and collaborators~\cite{cdr}.  
Each has strengths
and  weaknesses.  After the costed Technical Designs appear it will be
important to make a choice between  them which is driven by the quality of the
physics, not the nationality of the supporting laboratories.   Both designs
will initially be optimised for $\sqrt{s}=500 GeV$. TESLA at the moment appears
to  offer better luminosity for a given power consumption. Its larger aperture
and more widely spaced  bunches may make it easier to overcome alignment and
vibration problems.  Once a tunnel is built the ultimate energy of the machine
will be limited by the peak accelerating gradient.  The very best  performance
so far from TESLA-type cavities suggests that their present design, which is 
already capable of delivering the accelerating gradient needed for $\sqrt{s}=500 GeV$ in
the planned tunnel, might be developed to give maximum energy of a little 
over 800 GeV in the same tunnel.  The proponents
of the X-band design hope that with more and better klystrons they may eventually be 
able to raise the energy to 1 TeV.  The longer
term future, however, may rest with  
a change of technology to the CLIC concept developed
at CERN~\cite{clic}. This might eventually achieve six or  
seven times the 23.5 MV/m which
is the design gradient of TESLA for 500 GeV.  

The possibility to polarise of the  electrons to  
about 80\% is another legacy from SLC, and
the TESLA designers are hopeful that their  positron production scheme will
give 60\% polarisation, though R\&D is needed to prove this. 

Although the principal programme of the future collider will be with  $e^+ e^-$
collisions at high energies from the top  threshold at 350 
GeV upwards, there are other
important options being studied.  A Novosibirsk group~\cite{telginz} suggested
the Compton Collider, in which one or both beams from an $e^-  e^-$ linear
collider are intercepted with laser light a few millimetres before the
interaction point.  With   the right choice of laser energy, laser polarisation
and beam polarisation the resulting $\gamma  \gamma$  collisions can have more
than $1/10$ of the luminosity of the collider in its $e^+ e^-$ mode,  with a
peak at close to 80\% of the full energy.  In the $e^- \gamma$ mode the peak in the
photon energy  spectrum contrasts extremely favourably with the soft
bremsstrahlung of virtual photons used in $e \gamma$ scattering
until now.  There is R\&D to be done before the Compton Collider can be 
considered a proven option, but there are no known showstoppers~\cite{lbl}.  

No R\&D is needed to  prove the possibility of $e^- e^-$ collisions, just a
polarised electron gun at the nominally positron-end  of the linac.    The other
options which have been discussed in both Japanese and 
European studies are to modify the linac for  high luminosity
running at the $Z^0$ peak   and/or at the WW threshold.  

\section{Physics}

The main question to be tackled is ``why does the standard model work?''  How the
linear collider starts  to answer that will depend upon the second quesion,
``what will the LHC have discovered?''  The ECFA/DESY Linear Collider workshop at
Obernai~\cite{oberlhc} involved some of the leaders of the LHC physics studies
in  detailed debate on  the relative strengths of the two colliders in tackling
different analyses.  The conclusion was that there are a clear roles for both LHC and
the Linear Collider.  LHC may be first to find a Higgs boson (if we do  not see
it at LEP this year - and if it is indeed there to be seen), but the linear
collider is where its  properties will be pinned down and the kind of Higgs
boson established. 

\subsection{Properties of a light Higgs boson}
\label{mm}
\begin{figure}[htbp]
         \centering
         \epsfxsize=8cm
         \epsffile{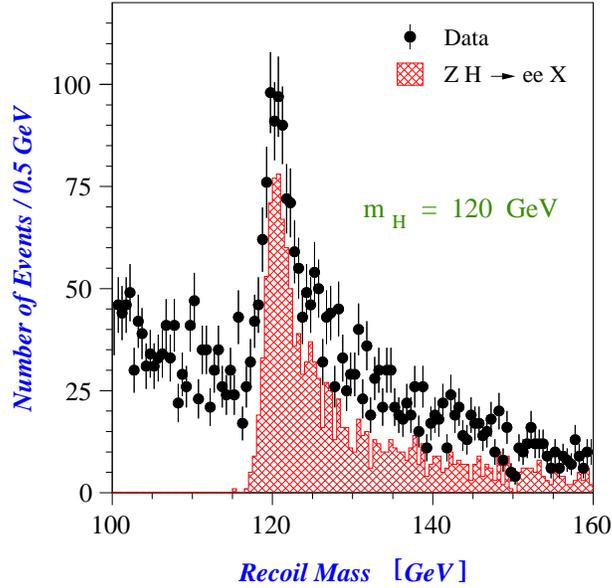}    
 \caption{\it
      Missing mass versus $Z^0 \rightarrow e^+ e^-$.
    \label{fig:higmm} }
\end{figure}
\begin{figure}[htbp]
         \centering
         \epsfxsize=8cm
         \epsffile{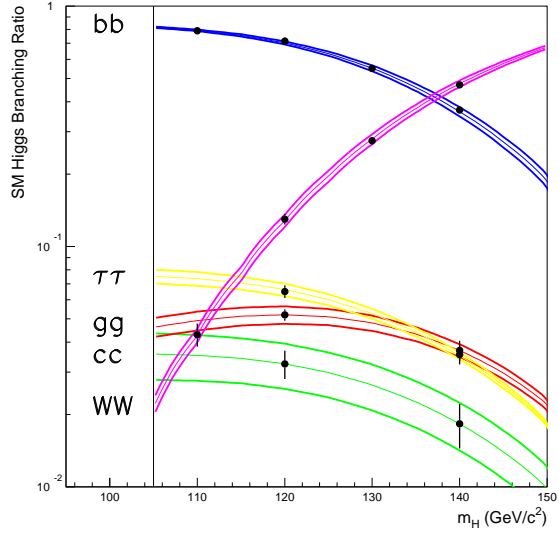}   
 \caption{\it Branching ratios for the standard model Higgs boson to the
 channels listed on the left ($b \bar{b}$ pairs, $\tau^+ \tau^-$, gluon pairs, 
 $c \bar{c}$ and $W^+ W^-$).  The error bars and bands represent
 estimates of the
 expected errors with $500 pb^{-1}$ of integrated luminosity at $\sqrt{s}= 500
 GeV$. 
      \label{fig:higerrs} }
\end{figure}

Figure~\ref{fig:higmm} shows the key to precision measurement of Higgs
boson properties at the  Linear Collider.  In the higgsstrahlung channel $e^+
e^- \rightarrow Z^0 H$, the $Z^0$ decays to $e^+  e^-$ and $\mu^+ \mu^-$can be
well measured and constrained to the $Z^0$ mass.  The missing mass  peak in the
recoiling system then contains all of the decays of the Higgs boson, visible or
invisible.  For  the standard model Higgs there should be no invisible
channels, and with a good microvertex detector it  will be possible to ``mine''
into
the peak in Figure~\ref{fig:higmm} for its branching ratios.  
And, of course, for the clearer channels such
as  $H \rightarrow b \bar{b}$, recoils against the much more copious hadronic
$Z^0$ decays can also be  used, not just the leptonic events shown 
in Figure~\ref{fig:higmm} which are used in
the search for exotic or invisible Higgs decays.  
See Figure~\ref{fig:higerrs} for the kind of
precision which may be achieved in different channels with an integrated
luminosity of $500pb^{-1}$ at $\sqrt{s}=500GeV$.  Note the large error bars on
the ``cc'' measurement.  The best possible vertex detector 
will be needed to resolve the 
charm-pair sample from the much  more copious beauty-pairs.
There are two promising strategies for measuring the total width of the
Higgs boson.  If its mass is  more than $115 GeV$, which it must be if LEP2 has
not seen it, then the total width can be found from  the ratio

\begin{equation}
\Gamma^{H}_{total} = \frac{\Gamma^{H}_{W^{*} W^{*}} 
B(H \rightarrow b \bar{b})}{B(H \rightarrow W 
W^{*} ) B(H \rightarrow b \bar{b})}.
\end{equation} 

Where the branching ratios on the bottom line can be measured in the
higgsstrahlung final states, and  the product on the top line is deduced from
the rate of the $W^{*} W^{*}$ fusion process $e^+ e^- \rightarrow  \nu \bar{\nu} H$ with
$H \rightarrow b \bar{b}$. 

If the Higgs mass is less than 140 GeV there is an alternative possibility:

\begin{equation}
\Gamma^{H}_{total} = \frac{\Gamma^{H}_{\gamma \gamma}B(H \rightarrow b \bar{b})}
{B(H \rightarrow \gamma \gamma) B(H \rightarrow  b \bar{b})}.
\end{equation} 

Again, the branching ratios on the bottom line may be measured from
higgsstrahlung final states  (though  $B(H \rightarrow \gamma \gamma)$ needs at
least $1000pb^{-1}$ integrated luminosity.  The top line
would have to be measured in the Compton Collider mode.  Perhaps  the strongest 
part of the case  for the Compton Collider will
come after a light Higgs boson has been discovered, because  $\Gamma^{H}_{\gamma
\gamma}$ receives contributions from any loops of new charged particles  which
couple perturbatively to the Higgs - so the Compton Collider could look ahead
to higher masses  than the Linear Collider would see directly.  The branching
ratio to gluons is also sensitive to loops which couple to colour,  
though the gluon pair final state
is likely to be hard to separate from $c \bar{c}$.

Measurement of the branching ratios shown in Figure~\ref{fig:higerrs} 
gives direct tests of
the Yukawa couplings of the  Higgs boson to different fermion 
flavours, which could
discriminate between SUSY models.  But the strongest of  the Yukawa couplings
of a light Higgs boson is to the top quark, and that cannot be measured in
Higgs  decays.  The Barcelona group~\cite{tth} has studied what can be done to
measure the $ttH$ coupling by  looking at the bremmsstrahlung of Higgs bosons
in top-quark pair production, $e^+ e^-  \rightarrow t \bar{t} H$,
which has a cross section of $2.5fb$ for 800 GeV collisions.  In 
$1000fb^{-1}$ of running they show that a signal can be found with expected
statistical errors of less  than 10 \%, but the backgrounds to be overcome are
huge, especially from top-pair production with  gluon radiation in the final
state.  Their verdict is that this coupling will only be measurable when the 
background calculations are all thoroughly understood, modelled and checked. 
This is one of the  studies which demands the best possible separation and
measurement of hadronic jets.     

\subsection{SUSY}

Another possible answer to ``why does the standard model work?'' could be that
supersymmetry exists at  an accessible scale.  If it does then there could be
many signatures of it within the range of the Linear  Collider - as well as the
possibility of 
measuring  non-SM coupling strengths for the light Higgs boson, as is likely to
be required if SUSY remains perturbative up to a high scale.  The slepton sector
will be hard to see at the  LHC, but any slepton states below the pair production
threshold will be directly accessible in $e^+ e^-$.   Feasibility studies show
that the masses of such states will be measureable to $ \pm 500 GeV$, using 
final state kinematics.  Threshold scans  could give even better precision. 
Chargino and neutralino pair production will also be much easier to  measure at
the Linear Collider.   The heavier Higgs bosons could be pair produced; $e^+
e^-  \rightarrow h A$, $e^+ e^- \rightarrow H^+ H^-$.  Polarisation will be
very important for their  analysis.  The Compton Collider  would be able to
prove that the A is CP odd.

\subsection{WW Physics}

\begin{figure}[htbp]
         \centering 
         \epsfxsize=13cm
         \epsffile{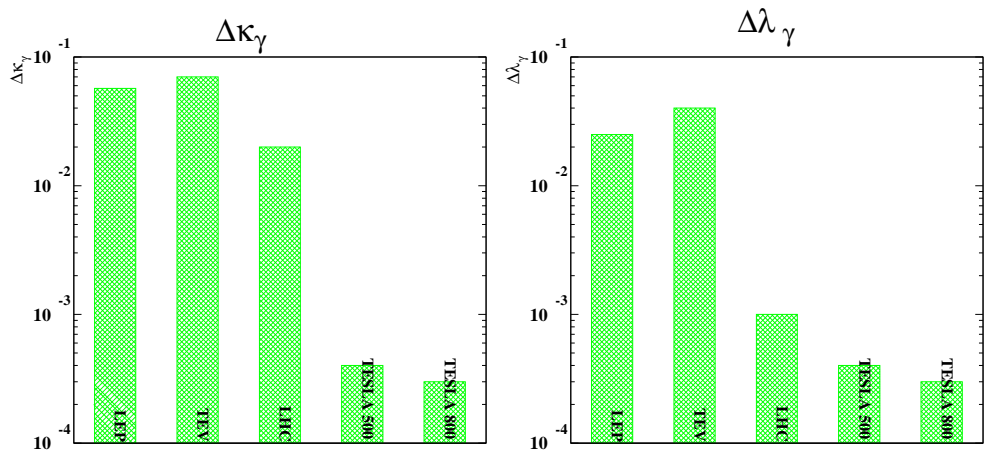}
 \caption{\it
      Estimated precision on anomalous triple gauge couplings at different
      colliders.
    \label{fig:tgc} }
\end{figure}

If no light higgs boson is found at LHC or Linear Collider, and no SUSY, then
the reason for the  standard model to work has to be some other  source of
Electroweak Symmetry Breaking which must  show up at some stage in  the
behaviour of the WW system.  The triple gauge boson couplings are  already
being investigated at LEP, but the Linear Collider will measure them with much
better  precision.  Figure~\ref{fig:tgc} shows that, for instance, 
the anomalous coupling $\Delta \kappa_\gamma$
can be determined to   $\sim 4\times 10^{-4}$ at a high luminosity 500 GeV
linear Collider, compared with $~\sim 6.10^{-2}$ at LEP or $~\sim 2.10^{-2}$ at
LHC.  Again, polarised beams will give extra sensitivity.  To exploit the  Linear
Collider  measurements it will be necessary to make substantial improvements to
the precision  of calculations of the electroweak matrix elements, including
higher order loops which have been  neglected so far.  To get such calculations
going in the ECFA/DESY Study we have just instituted a  dedicated theorists'
working group called the ``Loopverein''~\cite{loopverein}.

An important search channel for strong electroweak symmetry breaking will be direct
vector-vector  scattering, especially $W^+ W^-$.  The LHC will certainly be
able to look for this, but may be forced  to use only those events with a
leptonic W decay.  At 800 GeV the Linear Collier would have about  2000 $e^+
e^- \rightarrow W^+ W^-$ in 1000 $fb^{-1}$, all of which could be reconstructed
if the  detector has the good jet separation and energy flow resolution which
is planned in the detector design  from the ECFA/DESY Study.

\subsection{Other ways of pressing the Standard Model}

A classic task for the Linear Collider is to scan in small energy steps across
the threshold for top quark  pair production.  The rapid decay of the top quark
prevents the formation of narrow ``toponium''  resonances, but it also damps
higher order corrections to the shape of the threshold excitation.   Extensive
simulations in Europe, the USA and Japan have shown that the top quark mass can
be  determined from the scan to a precision of about $\pm 120 MeV$.  To  do
this it will be necessary to  limit the energy spread in the colliding beams to
less than about 2 parts in $10^3$.  The spectrum is also  spread out by
beamstrahllung - the radiation of energy from electrons before collision due to
scattering  off the coherent electromagnetic field of the opposing bunch - but
with current designs for the Collider  there is always a significant spike of
events in which the beamstrahlung loss is small.  These effects  combine to
give a  luminosity spectrum which can  be well monitored by using the
acollinearity of  Bhabha scattered electrons~\cite{frarymiller} in the endcap
region of the detector ($\sim 100$ to $\sim 300 mr$  from the beam direction).

With high luminosity and polarised beams it is also worth considering returning
to the LEP/SLC energy  region, running at the  $Z^0$ peak  to make precise
measurements of $sin^2 \theta_W$ (to $\pm  0.00002$). The $W^+ W^-$
threshold will also be worth revisiting, 
to reduce the error on the W mass to $~\pm 6MeV$.  This  will require
special by-pass facilities in the LINACs to maintain high luminosity; it is
inefficient to run  the beams through all of the accelerating the cavities at
much less than their maximum gradient.   TESLA should be able to produce $10^9$
$Z^0$s per year with a by-pass. Combining these precise  electroweak
measurements with the precise measurement of the  top mass will enable even
tighter  constraints to be placed on the parameters of the theory - probing
beyond the first Higgs boson, if one  has been found, or tying down the
possibilities for a Higgsless theory. 

There are also worthwhile jobs to be done in testing QCD.  High statistics at
500 GeV will allow  another point to be measured in the running of $\alpha_s$,
using event shapes and jet multiplicity  ratios, with comparable errors to the
LEP1 measurements~\cite{opalqcd}, and with a significantly 
increased lever-arm compared with
LEP2.  The Compton Collider in its $e^- \gamma$ mode will be ideally suited to
making  measurements of the photon structure function $F^{\gamma}_2(x,Q^2 )$ at
the highest possible  $Q^2$.  We are fighting hard to reach $Q^2 \simeq 1000
GeV^2$ at LEP2, with poor reconstruction of the Bjorken variable
 $x = Q^2 /(Q^2 + W^2)$ because the
target photon is drawn from a continuous bremmstrahlung  spectrum and the mass
$W$ of the $\gamma \gamma$ system can only be measured from the final state 
hadrons. At the Compton Collider with a nearly monochromatic photon beam $x$
will be much better  determined and there will be good statistics for $Q^2$ up
to 
$~\sim 10,000 GeV^2$.  This will allow a completely  independent measurement of the
strong coupling to be made from the evolution of the contribution due  to the
direct photon-quark interaction.

\section{Conclusions}
The designs for the Linear Collider machines are looking increasingly solid and
believable.  The choice between the options is about 2 years away. 
The physics programme of a linear collider will
be an essential complement to that of the LHC, whether or not a light Higgs boson is
discovered.  The detectors will have to be better than those at
LEP but they will be recognisably related to them and there are no very daunting
performance goals to be achieved.  More effort is needed to do feasibility
studies of physics processes and to study the detector and the machine-detector
interface.  Come to your local linear collider
worskhops~\cite{ecfad,charlie,acfa}.

\section{Acknowledgements}
I must thank all of the participants in the 2nd ECFA/DESY Study on Physics
and Detectors for a Linear $e^+ e^-$ Collider~\cite{ecfad} 
who have done most of the work reported in this talk.  In particular, Marco Battaglia and Ties
Behnke have supplied figures.

\end{document}